# LASER ION ACCELERATION TOWARD FUTURE ION BEAM CANCER THERAPY - NUMERICAL SIMULATION STUDY-

Shigeo Kawata[1,2], Takeshi Izumiyama[1], Toshihiro Nagashima[1], Masahiro Takano[1], Daisuke Barada[1,2], Qing Kong[3], Yan Jun Gu[3], Ping Xiao Wang[3], Yan Yun Ma[2,4], Wei Ming Wang[5]

1: Department of Advanced Interdisciplinary Sciences, Utsunomiya University, Japan
2: CORE (Center for Optical Research and Education), Utsunomiya University, Japan
3: Institute of Modern Physics, Fudan University, China
4: Department of Physics, National University of Defense Technology, China
5: Institute of Physics, Chinese Academy of Sciences, China

## ABSTRACT

**Background:** Ion beam has been used in cancer treatment, and has a unique preferable feature to deposit its main energy inside a human body so that cancer cell could be killed by the ion beam. However, conventional ion accelerator tends to be huge in its size and its cost. In this paper a future intense-laser ion accelerator is proposed to make the ion accelerator compact.
**Subjects and Methods:** An intense femtosecond pulsed laser was employed to accelerate ions. The issues in the laser ion accelerator include the energy efficiency from the laser to the ions, the ion beam collimation, the ion energy spectrum control, the ion beam bunching and the ion particle energy control. In the study particle computer simulations were performed to solve the issues, and each component was designed to control the ion beam quality.
**Results:** When an intense laser illuminates a target, electrons in the target are accelerated and leave from the target; temporarily a strong electric field is formed between the high-energy electrons and the target ions, and the target ions are accelerated. The energy efficiency from the laser to ions was improved by using a solid target with a fine sub-wavelength structure or by a near-critical density gas plasma. The ion beam collimation was realized by holes behind the solid target. The control of the ion energy spectrum and the ion particle energy, and the ion beam bunching were successfully realized by a multi-stage laser-target interaction.
**Conclusions:** The present study proposed a novel concept for a future compact laser ion accelerator, based on each component study required to control the ion beam quality and parameters.

## INTRODUCTION

Ion beams are useful for medical ion beam cancer therapy, basic particle physics, controlled nuclear fusion, high-energy sources, and so on.[1-10] Present conventional ion accelerators are huge in its size and its cost. On the other hand, a higher laser intensity has been realized by the chirped pulse amplification, and intense pulsed lasers are now available for experiments and applications. The energy of ions, which are accelerated by an interaction between the intense laser pulse and a gas target, reaches over a few tens of MeV.[11-25]

The issues in the laser ion acceleration include an ion beam collimation,[10,11,15] ion energy spectrum control, ion production efficiency,[24,25] etc. Depending on ion beam applications, the ion particle energy and the ion energy spectrum should be controlled. For example, ion beam cancer therapy needs 200~250MeV for the proton energy[26]. In recent researches ions are accelerated in an interaction of an intense laser pulse with a sold target or a near-critical density plasma.[11-25] However, the ion particle energy tends to be relatively low as shown above.

The paper shows first a novel concept of a future laser ion accelerator for ion beam cancer therapy. Then each component is discussed, and the components of the laser-based ion beam cancer therapy include ion sources by a solid target or a lower density gas plasma target, an ion post acceleration, an ion beam collimator and an ion beam buncher. We performed 2.5-dimensional particle-in-cell simulations[27] to investigate the laser ion acceleration. When an intense laser pulse propagates through the target plasma, it accelerates a part of target electrons. The electrons accelerated create an electric charge separation. The charge separation provides a strong electric field, for example, typically a few MV~10 MV/μm, by which the target ions are accelerated. This ion acceleration mechanism is called as TNSA (Target Normal Sheath Acceleration). In TNSA, the acceleration electric field is normal to the target surface. The TNSA is widely used for the ion source. However, the target deformation and the edge effect of the acceleration electric field induce the ion beam divergence transversely. Therefore, a collimation device is required. In the paper we present a solid target, which has holes behind the target for the collimation. The holes behind the target suppressed the source protons' divergence by reducing the ion source edge field. In addition, the target high-energy electrons also formed a high current and generate the azimuthal magnetic field in a plasma target, for example, ~KT. In the laser plasma interaction, the ion dynamics was affected directly by the electric field and the behavior of the electrons.[16-18] In the increase phase of the azimuthal magnetic field the inductive strong electric field was generated by the Faraday low. The ions were also accelerated by the inductive electric field. For each component of the laser ion accelerator, the TNSA and inductive electric fields play the important role to control the ion beam quality.

A novel concept was presented in the paper for a future laser ion accelerator, which should have an ion source, ion collimator, ion beam buncher and ion post acceleration devices. Based on the laser ion accelerator components, the ion particle energy and the ion energy spectrum are controlled; a future compact laser ion accelerator could be designed and realized for ion beam cancer therapy. In the post acceleration, a dramatic boost of ion particle energy and the energy spectrum control were realized in a laser plasma interaction; a few hundreds of MeV of the proton beam energy was successfully achieved by several times of the ion post-accelerations in the laser-plasma interaction. In addition, a mono-energetic proton beam was also produced.

Based on these research results on the laser ion accelerator components, the new concept for the laser ion accelerator becomes realistic for the future laser ion beam cancer therapy. The compact laser ion accelerator would provide a future daily-use ion accelerator for ion beam cancer therapy.

**A CONCEPT OF FUTURE LASER ION ACCLERATOR**

**Figure 1** shows a concept proposed for a future laser ion accelerator toward cancer therapy. In an intense laser interaction with a target, first protons are generated. The protons are accelerated by the strong electric field produced at the target by the laser-target interaction, in which the target electrons are expelled quickly and the target becomes a plasma. The high-energy electrons move around the target and the target ions stay rest during the short period of ~fs (= $10^{-15}$s). Between the ions and the high-energy electrons, a strong electric field is generated and accelerates protons gradually. In the ion source stage the ion beam tends to have a broad energy spectrum and a transverse divergence. For ion beam cancer therapy, the ion energy spectrum should be mono-energetic and the ion energy should be about 200-250MeV to deposit the ion energy in cancer cells inside human body.

Therefore, the laser ion accelerator would need post-acceleration devices to enhance the ion energy and collimators to suppress the ion divergence, as well as beam bunchers to compress the ion beam longitudinally. Depending on the requirements for the proton energy and the beam radius, the laser ion accelerator would be designed appropriately, as presented in **Fig. 1**.

In the next chapter each component consisted of the laser ion accelerator in **Fig. 1** is discussed and presented. The ion sources are first discussed, and then the post acceleration, the collimator and the buncher are presented.

# COMPONENTS OF LASER ION ACCELERATORS

## Proton beam source by a structured solid foil target
One of issues in the laser-ion acceleration is the energy conversion efficiency from laser to ions.

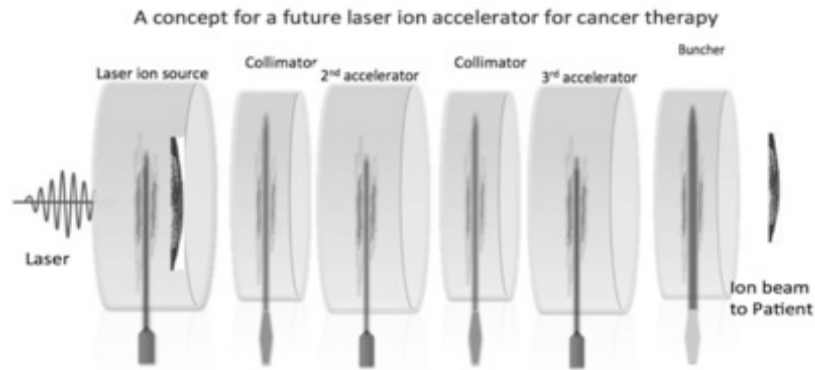

**Fig. 1**: Concept of an example future laser ion accelerator. For ion beam cancer therapy, proton energy must be 200-250 MeV. Therefore, further post-acceleration devices would be required to enhance the proton energy to achieve 200-250MeV of proton energy and also to control the ion energy spectrum. In addition, ion collimators are also needed to suppress the ion beam divergence. In order to compress the ion beam longitudinally, a beam buncher may be required. All the components are realized by laser target interactions.

When an intense short-pulse laser illuminates a thin foil, electrons in the foil obtain the laser energy and move around the thin foil. A part of the electrons placed at the surface irradiated by the laser is accelerated. The electrons form a high current and generate a magnetic field[14, 15]. The electrons form a strong electric field, and the ions are accelerated by the electric field. The laser transfers its energy mainly to the electrons near the thin foil surface. The target surface reflects a significant part of the laser energy. The energy conversion efficiency from laser to ions tends to be low. In this subsection, we employ an intense short-pulse laser and the double-layer hole target which consists of hydrogen and aluminum in 2.5-dimensional particle-in-cell simulations (see **Fig. 2(a)** and **Fig. 2(b)**). In an actual use a hydrogen rich material would be used for the hydrogen source. The reason why we employ aluminum is to prevent the target deformation. Aluminum ions are heavy compared with hydrogen, and the aluminum supplies electrons more than hydrogen. The maximum energy of the protons was about a few MeV in the plain target. The multihole target in **Fig. 2(b)** demonstrated significant increases in the energy conversion efficiency and the maximum energy. The holes, transpiercing the foil target, helped drastically to increase the laser energy absorption.[23, 24] The idea for the efficient laser ion acceleration by the fine structure was also proved by a recent experiment by D. Margarone, et al.[25]

Figure 3 shows a conceptual diagram of the multihole target. We employed a linear density gradient in 0.5λ to the target at the laser illumination surface to include a pre-pulse effect of the laser. The laser intensity was $I = 1.0 \times 10^{20}$ W/cm$^2$, the laser spot diameter was 4.0λ, and the pulse duration was 20fs. The laser transverse profile was in the Gaussian distribution. The laser wavelength was λ = 1.053μm. The calculation area was 40λ in the longitudinal direction and 30λ in the transverse direction. The free boundary condition was employed so that the boundaries do not reflect particles and waves. The ionization degree of the Al layer was 11. The initial Al target peak density was the solid one ($n_i = 42n_c$) and the H layer density was flat and $42n_c$. The initial particle temperatures were 1keV. Here $n_c$ stands for the critical density, at which the plasma electron frequency is identical to the laser frequency. The hole-diameter was 0.5λ.

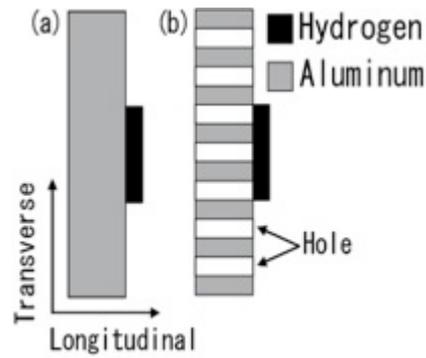

**Fig. 2**: Thin-foil targets: (a) a plain target and (b) a multihole target. An intense short-pulse laser gives electrons its energy, and the hot electrons are accelerated. The electrons form a strong electric field, and the protons are accelerated. The target surface reflects the laser. In the multihole target, the holes transpiercing the target help to enhance the laser-proton energy conversion efficiency.

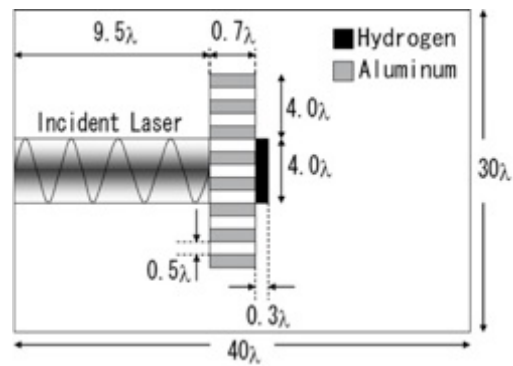

**Fig. 3**: A target structure used in this paper. We employ a double layer target which consists of Al and H. The Al layer has a linear density gradient in 0.5λ. The calculation area is 40λ in the longitudinal direction and 30λ in the transverse direction. There are twelve holes in the multihole target. The hole diameter is 0.5λ, and the hole distance is also 0.5λ.

First, we demonstrate the improvement of the energy conversion efficiency from laser to protons by the multihole target (**Fig. 2(b)**). **Figure 4** shows the distributions of the proton energy versus the longitudinal direction for both the cases of **(a)** the plain target and **(b)** the multihole target at 500fs. In **Fig. 4(a)** and **(b)** the protons in the multihole target were accelerated in the longitudinal direction more than those in the plain target. The maximum energy of the protons was 3.14MeV in the plain target and 10.0MeV in the multihole target. **Figure 5** shows the total-energy histories of the protons and electrons in both the cases. The peak of the laser intensity irradiated the target surface at about 55fs, and the total energy of the electrons reached the peak at about 60fs. In **Fig. 5**, the total energy of the protons reached about $5.30 \times 10^2$ J/m in the plain target and $3.10 \times 10^3$ J/m in the multihole target at 500fs. The energy conversion efficiencies to the protons were 1.50% in the plain target and 9.30% in the multihole target. The electric field formed in the multihole target was strong compared with that in the plain target, and the maximum value of the electric fields was 13.1MV/μm for the plain target and 27.5MV/μm for the multihole target at 60fs. **Figure 6** shows the histories of the total number $\Delta N/N$ of the protons accelerated over 0.1MeV in both the cases; the accelerated-proton number is normalized by the total proton number $N$ of the proton source. In **Fig. 6** the proton number in the multihole target reached about 2.7 times as many as that in the plain target at 500fs. In the multihole target, the strong electric field accelerated the protons significantly and contributed to the increase in the accelerated proton number. The electric field strength was about twice larger than that of the plain target. Therefore, the total-proton energy reached about six times as much as that in the plain target.

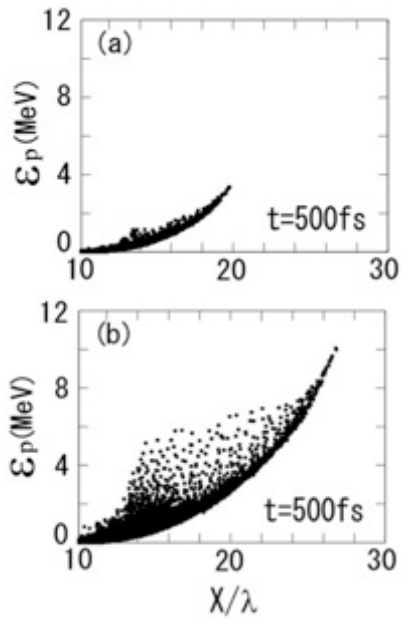 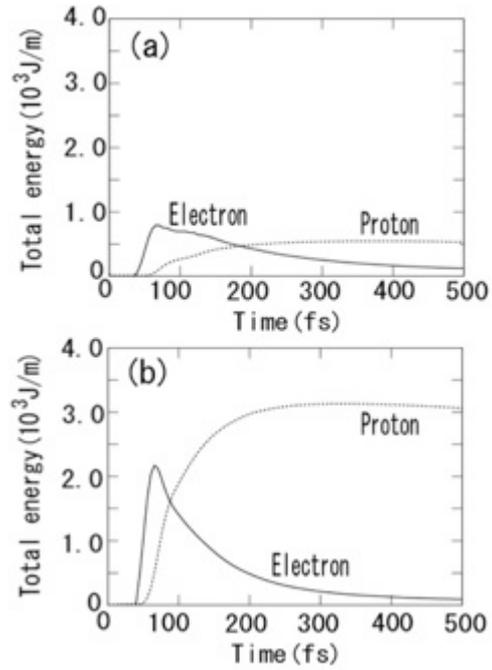

**Fig. 4**: Distributions of the proton kinetic energy at 500fs: (a) the plain target and (b) the multihole target. The proton kinetic energies become high significantly in the multihole target.

**Fig. 5**: Total-energy histories of the protons and electrons (a) in the plain target and (b) in the multihole target. The solid lines and the dotted lines are the histories of the electrons and the protons, respectively.

**Proton beam source by a gas plasma target**
Another promising candidate of the ion source target is a lower-density gas plasma, which can be created repetitively by a gas nozzle. In this simulation study, the hydrogen plasma target was located in $9.5\lambda < X < 29.5\lambda$ and $11.0\lambda < Y < 39.0\lambda$ in the simulation box. The target density was $0.7n_c$, and the edge density had a linear gradient scale from $0n_c$ to the maximum density of $0.7n_c$ in the scale of $2\lambda$ in the $X$ and $Y$ directions at the target edges. The laser intensity was $I=1.0 \times 10^{20}$ W/cm$^2$. The laser spot diameter was $10.0\lambda$, the laser was focused on the left edge of each target, and the pulse duration was 40fs. The laser transverse profile was in the Gaussian distribution, and the laser temporal profile was also Gaussian. The laser wavelength was $\lambda=1.05\mu$m. The simulation box was $80\lambda$ in the longitudinal direction and $50\lambda$ in the transverse direction.

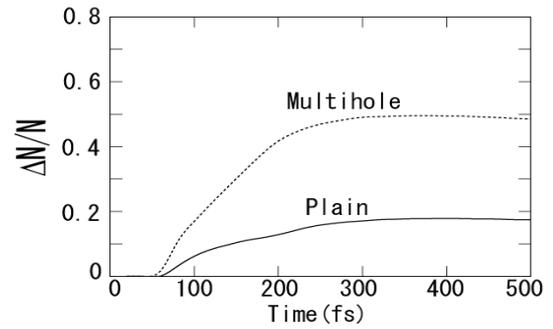

Fig. 6: Histories of the total number of the protons over 0.1MeV in the plain target and the multihole target. The solid line shows the history in the plain target and the dotted line shows the history in the multihole target. The proton number in the multihole target becomes about 2.7 times large compared with that in the plain target at 500fs.

The protons were accelerated by TNSA (Target Normal Sheath Acceleration) mechanism and the dipole vortex one[16] at the linear density gradient behind the target surface. **Figure 7** shows the longitudinal electric fields, which contributed to the ion acceleration at (a)$t$=130fs and (b)180fs, and the magnetic fields at (c)$t$=130fs and (d)190fs. The maximum acceleration electric field reached 16.7MV/$\mu$m at the end of the target area, and the maximum magnetic field reached 37.0kT in the 1st acceleration. The maximum proton energy was 38.9MeV at 700fs, and the energy convergence efficiency was 35.5% from the laser to the accelerated (>20MeV) protons.

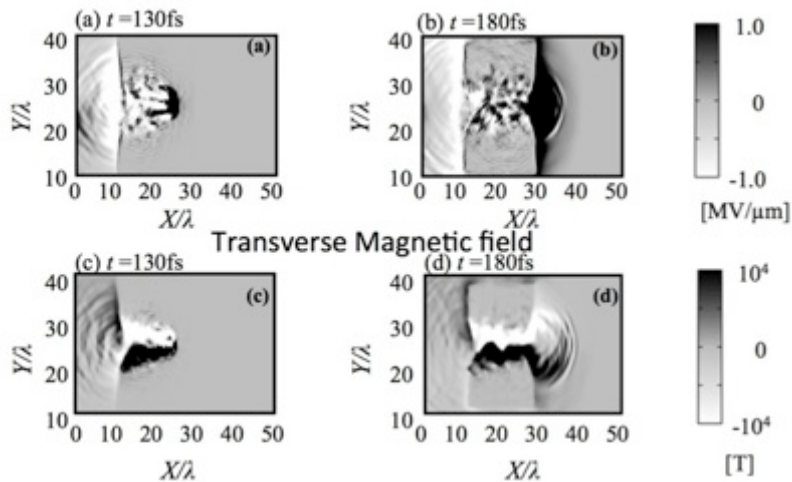

Fig. 7: The intense laser interacts with the hydrogen gas plasma, and protons are accelerated at the target rear surface by TNSA and the magnetic vortex acceleration. The longitudinal electric fields, which contribute the ion acceleration at (a)$t$=130fs and (b)180fs, and the magnetic fields at (c)$t$=130fs and (d)190fs. At the same time, the laser generates the high-energy electron inside of the target. A transverse magnetic field is also formed along the channel in the laser plasma interaction. During the increase phase of the magnetic field the inductive longitudinal electric field is created.

**Ion beam collimation by a structured solid target**

The laser-based ion accelerator also needs a collimation device, which reduces the ion beam divergence transversely.[10, 11, 15] The ion beam, generated by the laser ion source discussed above, has a small transverse velocity due the plasma target deformation and also the beam self charge. The collimation device reduces the ion transverse velocity to collimate the ion beam. Therefore, the collimation devices would be installed in the laser ion accelerator as shown in **Fig. 1**. In this subsection, a structured target shown in **Fig. 8** was employed. The proton beam obtained in **Fig. 4(b)** was introduced to the collimation device. The fine structure of the target left layer in **Fig. 8** absorbed the laser energy efficiently, and generated high-energy electrons. The electrons moved around the target, and created the strong electric field normally to the target surface. At the right hand side of the collimation target there was a larger scale structure, at which the TNSA field was created. The transverse electric field reduced the proton transverse divergence. In this case, the laser intensity was $5 \times 10^{19}$ W/cm$^2$, the pulse length was 100fs, and the spot size was 30λ. **Figure 9** shows the electric field distribution. **Figure 10** presents the normalized proton beam divergence distributions for the original input proton beam and for the collimated proton beam by the transverse electric field. The proton beam was collimated successfully by the collimation device in **Fig. 8**.

In addition to the collimation device in **Fig. 8**, S. Lund and his colleagues have proposed another collimation device using thin foils[28]. The thin foils prevent the electron return current and also shields the ion beam self charge by the mirror charge so that the ion beam self magnetic field contributes to the ion beam collimation.

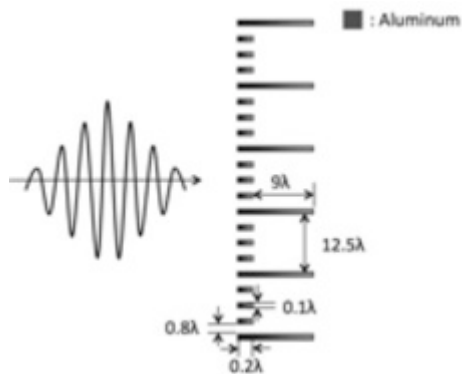

**Fig. 8**: A collimation device for ion beam. The Al structured target is illuminated by an intense laser. The fine structure absorbs the laser energy efficiently, and generates high-energy electrons. The electrons move around the target, and at the right hand side the electric field is created normally to the target surface. The transverse field is generated by the electrons and collimates the proton beam.

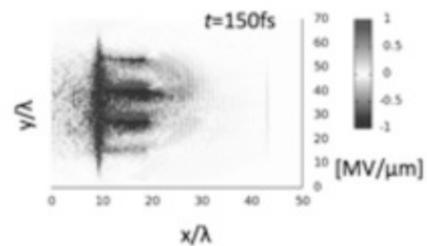

**Fig. 9**: The transverse electric field is successfully generated, and reduces the proton divergence.

**Ion beam bunching device**

In order to reduce the proton energy divergence and to reduce the ion beam longitudinal length, a bunching device would be required. As discussed above, when a structured solid target (see **Fig. 11**) is illuminated by an intense laser, a strong electric field is generated at the target rear side. The electric field contributes to the ion beam tail acceleration to reduce the ion beam energy divergence. In this subsection, the laser intensity was $2 \times 10^{20}$W/cm$^2$, the laser spot size $10\lambda$, and the laser pulse length 160fs. The structured Al target was employed (see **Fig. 11**).

**Figure 12** shows the results for the proton beam bunching. Initially the proton beam had the velocity divergence as shown in **Fig. 12(a)**. If the buncher was not used, the proton beam was elongated as shown in **Fig. 12(c)**. However, when the buncher was employed, the proton beam velocity divergence was reduced successfully and the proton beam length was kept short as shown in **Fig. 12(b)**.

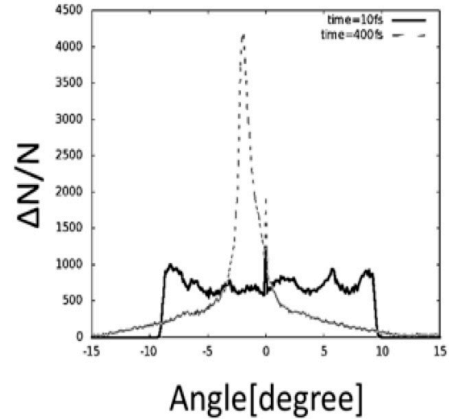

Fig. 10: Normalized divergence angle distributions for the original proton beam and for the collimated beam. The collimation device reduces the proton transverse divergence successfully.

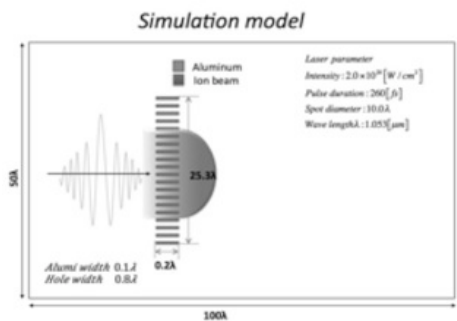

Fig. 11: A bunching device illuminated by an intense laser. The structured Al target absorbs the laser efficiently, and a strong electric field is generated to accelerate the ion beam tail in this specific case.

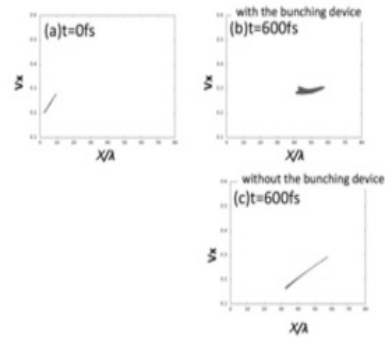

Fig. 12: The pre-accelerated proton beam is introduced to the bunching field. At $t=0$ the proton beam has the velocity divergence in **(a)**. With the longitudinal velocity divergence, the proton beam is elongated (see **(c)**). When the bunching device is used, the proton beam velocity divergence is reduced well as shown in **(b)**.

**Post-acceleration device**

By the ion beam source based on the laser target interaction, protons were accelerated to a few MeV ~ 10MeV with the laser intensity of the order of $10^{20}$W/cm$^2$. For the proton cancer therapy, the proton energy should reach about 200~250MeV. In addition, the proton energy spectrum should be also controlled well. Therefore, an additional acceleration and a method for the energy spectrum control are required for the ion beam cancer therapy. The result in **Fig. 7** also showed that the gas plasma target would provide a promising post-acceleration device.

When the intense laser pulse propagates through the plasma, it accelerates a part of electrons. The electrons form a high current and generate the azimuthal magnetic field around the laser axis. The electrons form the strong magnetic field; in the increase phase of the azimuthal magnetic field, the strong inductive electric field was generated by the Faraday low. The ions were accelerated by the inductive electric field inside the plasma, as well as the TNSA (Target Normal Sheath Acceleration) mechanism and the dipole vortex one[16].

In this subsection, a multi-stage laser ion post-acceleration (see **Fig. 13**) was investigated for the control of the ion energy and the ion energy spectrum. The identical laser and the target were employed with those used in **Fig. 7**.

In the 1st acceleration stage, first the protons were accelerated by TNSA mechanism and the dipole vortex one behind the target surface. Secondly, the high-energy ions propagated to the next 2nd post-acceleration. The 3rd and 4th ion post-accelerations were continued. In the 1st and 2nd acceleration stages, the beam ions were mainly accelerated by the TNSA acceleration and dipole vortex acceleration mechanisms. However, in the 3rd and 4th post-accelerations the ions were accelerated by the inductive acceleration based on the Faraday law inside the plasma target as well as the TNSA and dipole vortex acceleration mechanisms behind the target surface.

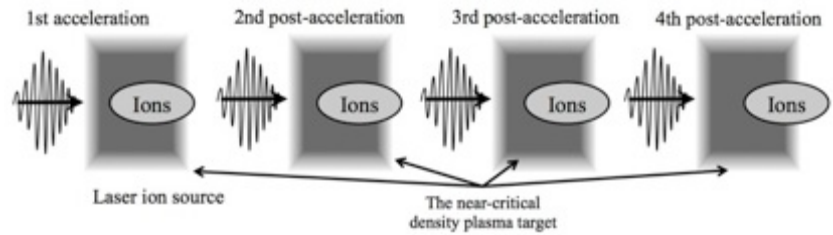

**Fig. 13**: The conceptual diagram for the post-acceleration in the laser plasma interaction. The generated ions from laser ion source are accelerated by several-stages post-acceleration. We employ the near-critical density plasma target, which consists of hydrogen. In this example, the four-stages ion acceleration is performed.

When an intense short pulse laser illuminates the near-critical density plasma, the inductive acceleration field moves with a speed $v_g$, which is less than $c$ depending on the plasma density: $v_g = c\sqrt{1 - \omega_{pe}^2/\omega^2}$, where $c$ is the speed of light, $\omega_{pe}$ the electron plasma frequency and $\omega$ the laser frequency. Additionally, the beam protons had a higher speed especially in the 4th stage, compared with that in the 1st and 2nd stages. Therefore, the inductively-accelerated ions were kept accelerating for a long time inside the near-critical density plasma target. The distance between the two adjacent targets was $60\lambda$.

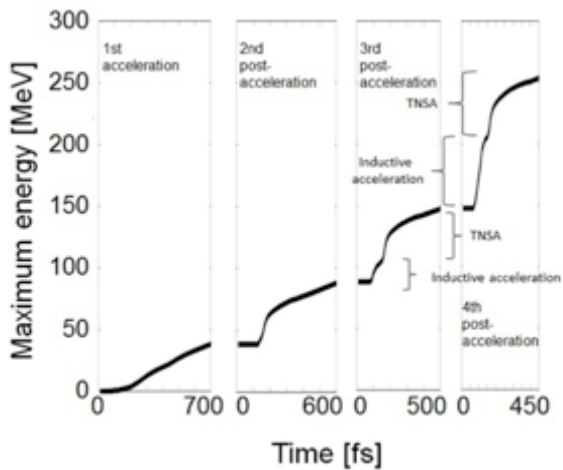

**Fig. 14**: The histories of the maximum proton energy from the 1st acceleration to the 4th post-acceleration. The maximum proton energy is remarkably accelerated by the four-stages acceleration. The maximum proton energy is finally about 254.0MeV in the 4th post-acceleration at 450fs.

**Figure 14** shows the histories of the maximum proton energy from the 1st acceleration to the 4th post-acceleration.

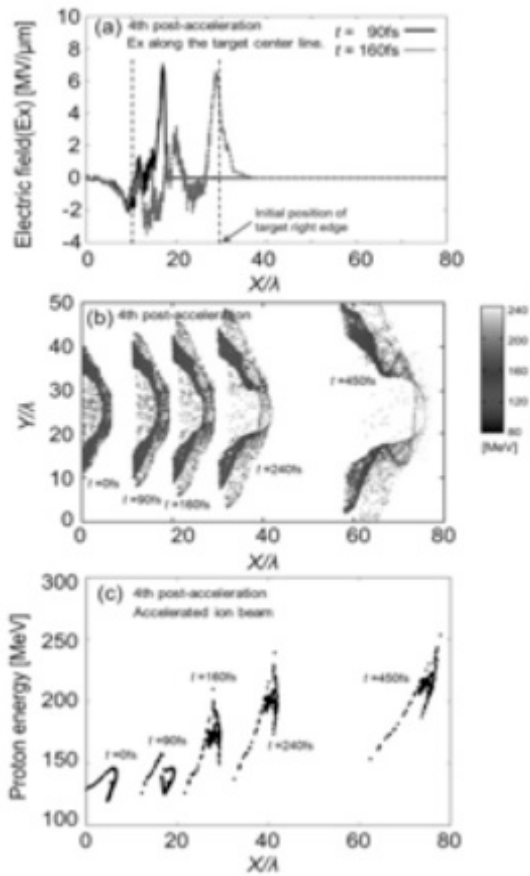

The maximum proton energy was 38.9MeV at the 1st acceleration at 700fs, 89.9MeV at the 2nd post-acceleration at 600fs and 149MeV at the 3rd post-acceleration at 500fs. Finally the maximum proton energy reached 254MeV in the 4th ion post-acceleration at 450fs.

Especially, in the 4th stage the proton energy was about 150~250MeV during the inductive acceleration phase (**Fig. 14**), and the proton speed was about $0.566c \sim 0.653c$, which corresponded to $v_g \sim 0.548c$ (the speed of laser in the plasma density of $0.7n_c$). Therefore, the ions were accelerated continuously by the inductive field in the target and then accelerated further by the TNSA and the magnetic vortex mechanisms successfully.

**Figure 15(a)** shows the acceleration electric field $Ex$ along the center line of target in the longitudinal direction. The black solid sine is $Ex$ at $t$=90fs and the grey dotted line is $Ex$ at $t$=160fs for the 4th post-acceleration in **Fig. 15(a)**. The inductive electric field for the ion acceleration was generated inside the near-critical density plasma target (solid line in **Fig. 15(a)**). The acceleration TNSA electric field was generated behind the target (dotted line in **Fig. 15(a)**). **Figure 15(b)** shows the proton spatial distributions and **Fig. 15(c)** shows the proton energy distributions in $20.0\lambda < Y < 30.0\lambda$ at $t$=0fs, 90fs, 160fs, 240fs and 450fs. The proton energy reached 254MeV at $t$=450fs. In this subsection the laser spot size was $10\lambda$, and as shown in **Fig. 15(b)** the core protons accelerated are almost located in

**Fig. 15**: Acceleration electric fields averaged over the laser one cycle along the center line of the plasma target in the longitudinal direction. The black solid line is $Ex$ at $t$=90fs and the grey dotted line is at $t$=160fs in the 4th post-acceleration. (b) The proton spatial distributions at $t$=0fs, 90fs, 160fs, 240fs and 450fs. The color shows the energy of the protons. (c) The energy distributions at $t$=0fs, 90fs, 160fs, 240fs and 450fs for the protons existing in $20.0\lambda < Y < 30.0\lambda$ at 450fs. The maximum proton energy reaches to finally 254MeV in the 4th post-acceleration at 450fs.

$20.0\lambda < Y < 30.0\lambda$.

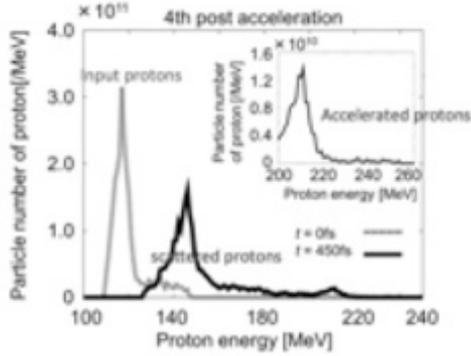

**Fig. 16**: Energy spectra of protons in the 4th post-acceleration. The dotted lines are the input ion beam to the 4$^{th}$ stage, and the solid lines are the output ion beam. Figures present all the ions, including the scattered ions transversely, and the inset figure shows the high-energy core part of the beam ions located in 20.0$\lambda$ < $Y$ < 30.0$\lambda$. The core part of the beam ion is useful for practical purposes.

**Figure 16** shows the ion beam energy spectra for the 4th post-acceleration. The dotted line shows the input protons to the 4th stage, and the solid line shows the output protons from the 4th stage. **Figure 16** present all the ions, including the scattered ions transversely (see **Fig. 15(b)**). The inset figures in **Fig. 16** show the high-energy core part of the beam ions located in 20.0$\lambda$ < $Y$ < 30.0$\lambda$, and the core part of the beam ions could be useful for practical purposes. The inset figure in **Fig. 16** shows a quasi mono-energetic spectrum, which was also realized in the multi-stage ion acceleration.

In the present case, our simulation parameter study showed that the synchronization timing requirement $\Delta t$ between the laser pulse and the ion beam wass about 10 fs so that the decrease in the ion maximal energy was limited by 10.2%. The TNSA field and the inductive acceleration field were generated by the laser pulse, whose pulse duration time was 40fs in this subsection. The laser pulse risetime $T_{Lr}$ was ~20fs. Especially the risetime $T_{Lr}$ defined the duration time for the inductive acceleration field in the plasmas. Therefore, the synchronization requirement should be less than the laser risetime: $\Delta t << T_{Lr}$.

## CONCLUSIONS

In this paper we have proposed and discussed the concept for a future laser ion accelerator for cancer therapy. The practical laser ion accelerator would consist of an ion source, ion beam collimators, ion beam bunchers and post-accelerators depending on the requirements for the ion particle energy, the ion energy spectrum, the beam radius and the ion beam pulse length. The each realistic component was also proposed and studied in this paper. In the laser ion acceleration, ions were accelerated by the strong electric field, generated by the interaction of the intense laser with the target. The electric field was created by the TNSA field and the inductive electric field. The acceleration field gradient was rather large (>10GV/m) compared with that (<100MV/m) of the conventional accelerator. Based on the scientific results presented in this paper, the realistic laser ion accelerator could be designed as shown in **Fig. 1**. Each component and combinations of the components provide a high controllability of the ion beam generated by the laser ion accelerator to meet variable requirements for ion beam cancer therapy.

The ion sources proposed in this paper realized a high energy convergence efficiency more than a few tens percent by using the fine structure solid target (**Fig. 3**) or using a lower-density gas target (**Fig. 7**). In order to control the ion energy and the ion energy spectrum, the gas target provided the efficient inductive post-acceleration device. The proton energy of 200~250 MV was achieved and the mono-energetic energy spectrum was also realized. Laser produced ion beams tend to have a transverse divergence. For the ion beam collimation the structured target (**Fig. 8**) was also proposed to collimate the proton beam. S. Lund, etc. also proposed the thin foils based new collimator[27]. The ion beam bunching device (**Fig. 11**) was also proposed to control the ion beam length depending on the therapy requirement.

Conventional ion accelerators for the ion beam cancer therapy tend to be a large size and to be expensive. The laser ion accelerator proposed in this paper may open a new method to produce a controlled ion beam for the cancer therapy in its smaller size and in its reasonable cost.

## DISCUSSIONS

The novel laser ion accelerator was proposed toward a future ion beam cancer therapy, and the laser accelerator proposed provides a flexible capability to adjust the requirements. However, there are some issues remained to be studied further: for example, the laser ion accelerator would need multiple laser pulses with different focusing points with a high repetition rate, say 1~10 Hz or so to kill the cancer cells. The repetitive operation of the intense laser system might be available even today at the intensity of the order of $10^{20}$ W/cm$^2$ or more. So the repetitive laser operation may need further studies. The present efficiency of the laser pulse emission itself is rather low, almost less than 1 percent. Another issue is how to prepare and install the targets in the repetitive operation environment. A gas jet nozzle would work well to produce the gas target repetitively employed in this paper. In addition, the structured thin foil target could be prepared in a tape form for the repetitive usage. For the ion source and for the post acceleration device, the laser based acceleration and post-acceleration provide a high acceleration gradient, and they contribute to shorten the accelerator devices. The conventional collimation and bunching devices could also substitute the laser-based collimators and bunchers, depending on the required quality for the ion beam. These issues should be studied in the near future to obtain the real laser ion accelerator for the cancer therapy.

## ACKNOWLEDGEMENTS

This work was partly supported by MEXT, JSPS, ASHULA project/ILE/Osaka Univ. and CORE (Center for Optical Research and Education, Utsunomiya Univ., Japan). The authors would also like to express their appreciations to Prof. Z. M. Sheng, Prof. J. Limpouch, Prof. O. Klimo and Prof. A. Andreev for their fruitful discussions on the subject. The authors would also like to extend their acknowledgements to Prof. Toshio Ohshiro, Prof. Cheng-Jen Chang, Prof. K. Ting and the organizers of APALMS2012 for their encouragement to present our research results at APALMS2012 and in Laser Therapy.